\documentclass[apj]{emulateapj}
\usepackage{gensymb}
\usepackage{amsmath}
\usepackage{graphicx}
\newcommand{\rhopdf}[0]{$\rho-$PDF}
\newcommand{\npdf}[0]{$\mathcal{N}$-PDF}
\begin{document}
\title{A Simple Perspective on the Mass-Area Relationship \\in Molecular Clouds}
\shorttitle{The Mass-Area Relationship}
\slugcomment{MNRAS in press}
\shortauthors{Beaumont et al.}
\author{Christopher N. Beaumont\altaffilmark{1,2}, Alyssa A. Goodman\altaffilmark{2}, Jo\~ao F. Alves\altaffilmark{3}, Marco Lombardi\altaffilmark{4}, Carlos G. Rom\'an-Z\'u\~niga\altaffilmark{5}, Jens Kauffmann\altaffilmark{6}, Charles J. Lada\altaffilmark{2}}
\altaffiltext{1}{Institute for Astronomy, University of Hawai'i at Manoa, 2680 Woodlawn Drive, Honoulu, HI 96822, USA}
\altaffiltext{2}{Harvard Smithsonian Center for Astrophysics, MS 42, 60 Garden Street, Cambridge, MA 02138, USA}
\altaffiltext{3}{Institute of Astronomy, University of Vienna, T\"urkenschanzstr. 17, 1180 Vienna, Austria}
\altaffiltext{4}{Universit\'{a} degli Studi di Milano, Dipartimento di Fisica via Celoria 16, I-20133 Milano, Italy}
\altaffiltext{5}{Instituto de Astronom\'{i}a, Universidad Nacional Aut\'{o}noma de M\'{e}xico Km 103 Carr. Tijuana-Ensenada, Ensenada BC 22860, Mexico}
\altaffiltext{6}{Jet Propulsion Laboratory, California Institute of Technology, 4800 Oak Grove Drive, Pasadena, CA 91109, USA}
\begin{abstract}
Despite over 30 years of study, the mass-area relationship within and among clouds is still poorly understood both observationally and theoretically. Modern extinction datasets should have sufficient resolution and dynamic range to characterize this relationship for nearby molecular clouds, although recent papers using extinction data seem to yield different interpretations regarding the nature and universality of this aspect of cloud structure. In this paper we try to unify these various results and interpretations by accounting for the different ways cloud properties are measured and analyzed. We interpret the mass-area relationship in terms of the column density distribution function and its possible variation within and among clouds. We quantitatively characterize regional variations in the column density PDF. We show that structures both within and among clouds possess the same degree of ``universality'', in that their PDF means do not systematically scale with structure size. Because of this, mass scales linearly with area.
\end{abstract}

\maketitle

\section{Introduction}
The hierarchical spatial-kinematic structure of molecular clouds provides clues about the processes that dictate cloud evolution, and sets the initial conditions for star formation. Cloud structure is influenced by several processes -- magneto-hydrodynamic turbulence, gravity, stellar feedback, internal and external radiation fields, etc. -- and is extremely complex, exhibiting features across a wide range of scales. Absent a framework for describing cloud structure in its entirety, researchers have applied a bevy of empirical techniques to characterize observed and simulated data; examples include the distribution functions for density, column density, velocity and mass, size-linewidth relationship, structure function, delta-variance relationship, principal components analysis, and spectral correlation function (see \citealt{Elmegreen04} and references therein). The relationships and interdependencies that exist between these various descriptors are rarely obvious.

Larson's scaling relationships \citep{Larson81} are among the most versatile empirical characterizations of molecular cloud structure. The third of these relationships (which we will refer to as L3) states that the average volume density of a molecular cloud scales inversely with its size or, alternately, that a cloud's mass is proportional to its area.  In the thirty years since its publication, Larson's mass-area relationship has been subject to numerous re-investigations. Table \ref{tab:larson3} lists several published measurements of mass-area relationships, assuming a functional form $(M/M_\odot) = \gamma ~(A/{\rm pc^2})^\beta$. These are all measurements of Galactic cores, clouds, and cloud complexes (for extragalactic measurements, see \citealt{Bolatto08, Hughes10}). There is substantial scatter among these values, especially in the constant of proportionality $\gamma$.

Much of the uncertainty about L3 (and the scatter in Table \ref{tab:larson3}) is due to the heterogeneous ways in which cloud properties are measured. The conversion from intensity to mass is non-trivial for most tracers; dust emission is influenced by uncertain temperature and emissivity variations, while molecular line emission is affected by opacity, chemistry, and changing excitation conditions \citep{Goodman09}. Several authors have suggested that L3 is an observational artifact, owing to the fact that individual surveys are sensitive to a relatively narrow range of surface brightnesses, artificially driving measurements towards the $M \propto A$ relationship \citep{Kegel89, Ballesteros02}. Furthermore, clouds do not have sharp edges that define a natural size; researchers usually use specific contours or moment measurements to describe a cloud's area. Finally, it is difficult to reliably segregate clouds located in the Galactic mid-plane. All of these factors will affect the mass-area relationship.

Larson's relationships are also studied in disparate regions. Indeed, the sample in \cite{Larson81} comprises small substructures within clouds, individual clouds, and large cloud complexes. It is important to distinguish cloud-to-cloud comparisons from comparisons of structures within clouds, as these two regimes are governed by different processes and may yield different scaling relationships.

Lombardi, Alves and Lada (\citeyear{Lombardi10}, hereafter LAL10) addressed many of these issues by measuring cloud sizes and masses in a systematic way. Using 2MASS-derived \citep{2mass} extinction maps (which do not suffer the same dynamic range limitations or calibration uncertainties inherent to dust and gas emission data; \citealt{Goodman09}), they measured the mass and area of 11 nearby clouds. All of these clouds are far enough out of the mid-plane that background confusion is low. They experimented with different cloud boundaries, using the extinction contours at $A_K = (0.1, 0.2, 0.5, 1.0, 1.5)$. Comparing different clouds at a constant extinction threshold, they found that the data obey $M \propto A$ to excellent approximation, although the constant of proportionality changes with each threshold (Figure \ref{fig:compare}a). Their results provide strong evidence that the correlation between the areas and masses of clouds is a real phenomenon, and not an observational bias. They also demonstrate that a lognormal column density distribution (PDF) can reproduce this relationship, as long as the cloud-to-cloud variation in the lognormal width parameter ($\sigma$) is not too large (they report a scatter of $\sim 20\%$). From their results, they conclude that molecular cloud structure is ``universal'', in the sense that different clouds possess similar column density PDFs, and as a result of this obey $M \propto A$.

\begin{figure}
\includegraphics[width=5in]{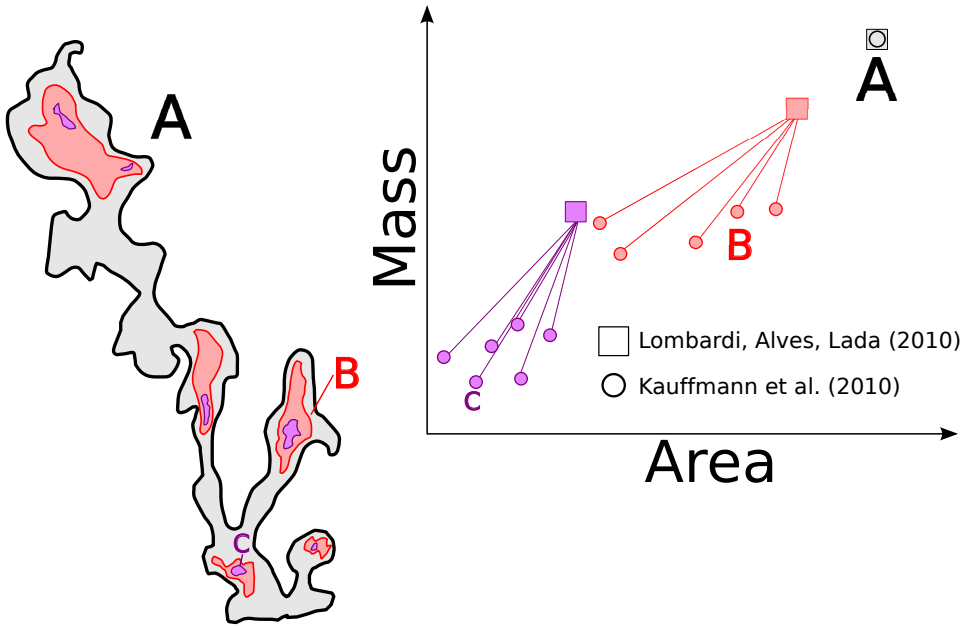}
\caption{Schematic comparison of how cloud masses/areas are measured by \cite{Lombardi10} and \cite{Kauffmann10a}. Each closed contour has a well-defined mass and area, as described above. \cite{Kauffmann10a} treat each closed contour separately (circles), whereas \cite{Lombardi10} add up the mass and area of every region defined by a given intensity threshold (squares). }
\label{fig:schematic_1}
\end{figure}

\cite{Kauffmann10a, Kauffmann10b} have also recently studied the mass-area relationship of nearby clouds using extinction data. Like LAL10, these studies defined cloud substructures by examining regions with extinctions above a given threshold. However, Kauffmann et al. analyzed each closed contour separately; Lombardi, Alves and Lada merged all regions above a given contour level when measuring cloud masses and sizes (see Figure \ref{fig:schematic_1}).  Superficially, the mass-size relationships that Kauffman et al. derived \textit{within} clouds seem to exhibit more heterogeneity than the relationships that LAL10 derived \textit{between} clouds (Figure \ref{fig:compare}b). This raises a question: to what extent is the ``universality'' in cloud structure emphasized by LAL10 (and attributed to the global similarity of cloud column density PDFs) challenged by the internal structural heterogeneity suggested by the work by Kauffmann et al.?

Answering this question requires more carefully comparing the methodological differences between these studies, and is the focus of this paper. We emphasize the distinction between the mass-size relationship between and within clouds, and interpret both in the context of the column density PDF and its variations. We first consider the inter-cloud relationship, and provide a few important critiques to LAL10 regarding the conditions under which a collection of column density PDFs will yield L3. These critiques lead to a more specific definition of ``universal'' cloud structure. We then consider to the intra-cloud mass-area relationship. This relationship can also be connected to the column density PDF, and in particular can diagnose variations in the PDF within the cloud. We demonstrate that the substructures within Perseus possess the same quality of universality as do structures across clouds.

\section{Data and Analysis}
In this paper we use the same extinction data presented by LAL10. These maps are produced using the NICEST technique of measuring the reddening of background stars due to foreground molecular clouds \citep{nicest}.  The maps are based on 2MASS data \citep{2mass}, have an effective angular resolution of $1-3'$, and a noise of $A_K \sim 0.06$ ($N \sim 5 \times 10^{20} \rm ~ cm^{-2}$) -- comparable to the median column density within molecular clouds. We refer the reader to \cite{Lombardi10} for more details about the data reduction process.

We define cloud structures via contours in the extinction maps. The area of each closed contour is obtained directly by multiplying the surface area of each pixel by the number of pixels within the region. Likewise, the mass is obtained by integrating the extinction within the region, and using the relationship ${\rm 1~ mag~ }A_K = 180 M_\odot \rm pc^{-2}$ \citep{Rieke85}. Many authors further define a cloud's radius via $R \equiv \sqrt{A / \pi}$. However, to avoid confusion when dealing with significantly non-circular regions, we restrict our focus to area measurements only.

As mentioned above, LAL10 merge together all pixels above a given threshold when measuring mass and size -- that is, at a given threshold, each cloud is described by a single mass and size (Figure \ref{fig:schematic_1}). \cite{Kauffmann10a}, on the other hand, treat each closed contour separately, and thus are sensitive to potential variations within a cloud.

\begin{figure}
\includegraphics[width=6in]{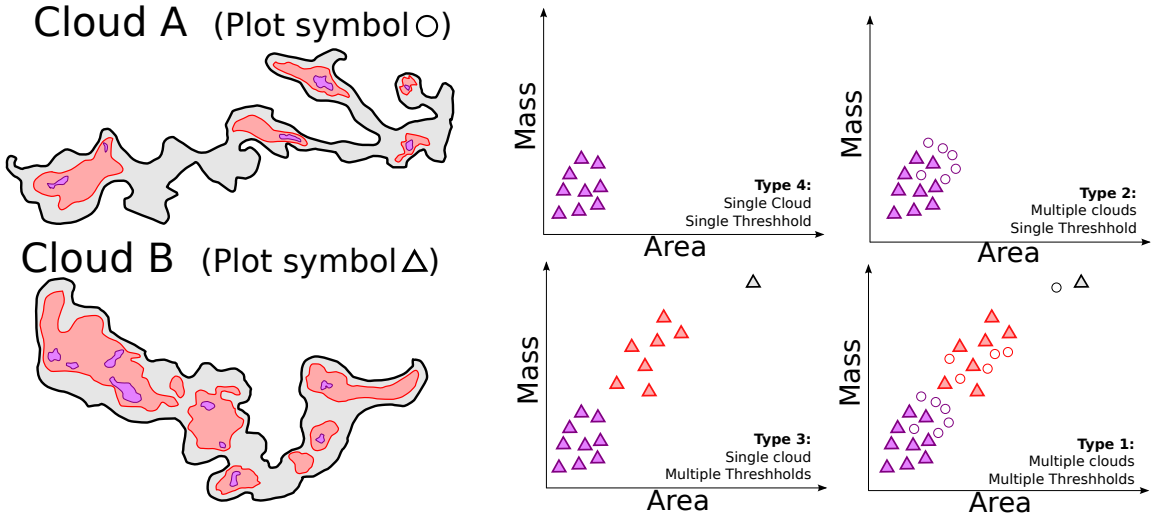}
\caption{The different possible mass-area relationships to measure. As in Figure \ref{fig:schematic_1}, each point corresponds to a mass and size measurement for a single closed contour in a cloud. The color of the point represents the contour value, and the shape represents the cloud. In this paper we focus on the relationships 2 and 4; that is, our comparisons of structures within and between clouds always adopt a fixed extinction threshold. The numbering scheme has been chosen to match the linewidth-size types defined by \cite{Goodman98}.}
\label{fig:schematic_2}
\end{figure}

Several mass-area relationships can be measured given a collection of clouds and their substructures. These are depicted schematically in Figure \ref{fig:schematic_2}, where each point corresponds to a mass and size measurement of a single closed contour. The shape of the point depicts which cloud each measurement comes from, and the color depicts the contour value used. Each relationship is defined by whether or not the comparison spans several clouds, and whether structures are defined at fixed or variable thresholds. Table \ref{tab:larson3} lists the comparison types used in the literature. We restrict our focus to comparisons of structures defined at a common extinction threshold (i.e. the Types 2 and 4 in the Figure). As \cite{Lombardi10} demonstrated, the choice of extinction threshold affects the offset in the mass-area relationship, since the same region defined using a lower extinction threshold will contain more area and mass. Furthermore, as both \cite{Lombardi10} and \cite{Kauffmann10a} note, comparing structures defined using different thresholds will bias the relationship, since structures defined at low thresholds sample systematically lower surface-density material.

\section{The inter-cloud mass-area relationship}
\label{sec:inter}
When cloud structures are identified and measured using extinction contours as described above, their masses and areas are related by
\begin{equation}
M \equiv \lambda \left<N\right> A
\label{eq:mna}
\end{equation}
where $\left<N\right>$ is the mean column density inside that contour, and $\lambda$ is a constant of proportionality that converts column density (particles per area) to surface density (mass per area). We stress that this equation does not necessarily imply that $M \propto A$, since the quantity $\left<N\right>$ can vary from region to region. As LAL10 note, the mean column density can also be expressed as an integral of the column density distribution function $P(x)$:
\begin{equation}
\left<N\right> \equiv \int_{N_0}^\infty x P(x) dx
\label{eq:truncated_mean}
\end{equation}
Here, $N_0$ is the column density threshold used to define the cloud boundary, and $P(x)$ is the normalized column density distribution function -- the units of the function are inverse column density, and its integral over all column densities is one. Importantly, $P(x)$ describes only the distribution of pixels within the region of interest -- this distribution can, in principle, vary from region to region. \cite{Lombardi10} circumvent any possible regional PDF variation by treating all pixels above $N_0$ as a single entity, even if this threshold divides the cloud into several disconnected regions (Figure \ref{fig:schematic_1}). In this way, the distribution function $P(x)$ used in Equation \ref{eq:truncated_mean} is always the same for a single cloud.

It is clear from Equation \ref{eq:mna} that Larson's mass-area relationship (i.e., $M \propto A$) is approximately satisfied whenever $\left<N\right>$ is approximately constant for the objects studied. Alternately, if a common threshold $N_0$ is used to define all objects in question, L3 is approximately satisfied whenever $P(x)$ is approximately constant across regions.

The clouds presented in \cite{Lombardi10} exhibit excellent agreement with L3. The authors further show that, if clouds possess log-normal column density PDFs, then the parameters describing the location and width of this distribution must be similar across regions to reproduce this result. The log-normal fits to these PDFs exhibit 60\% and 20\% scatter in the location ($\mu$) and width ($\sigma$) parameters, respectively. It is important to emphasize that the lognormal distribution is not the only distribution capable of reproducing L3 (indeed, any $P(x)$ can). Generalizing the analysis in LAL10 to arbitrary PDFs reveals interesting connections between the column density distribution and mass-area relationship, and more precisely defines under what conditions L3 holds.

Consider a collection of mass and area measurements $(M_i, A_i)$ for $q$ regions, using a common threshold $N_0$, and a power-law model:
\begin{equation}
\log M_i = a + b \times \log A_i
\label{eq:model}
\end{equation}

We seek the conditions under which the data are well described by L3 -- i.e., Equation \ref{eq:model} with $b=1$. Note that we can re-arrange Equation \ref{eq:mna} as follows:
\begin{equation}
\log M_i \equiv \log \lambda + \overline{\log \left< N\right>} + 1 \times \log A_i + \epsilon_i
\label{eq:actual}
\end{equation}
where $\overline{\log \left<N\right>}$ is the sample mean of the individual $\log \left<N\right>$ terms, and $\epsilon_i \equiv \log \left<N\right>_i - \overline{\log \left< N \right>}$. Equations \ref{eq:model} and \ref{eq:actual} have the same form, and differ by a residual term $\epsilon_i$. If the $\epsilon_i$ terms are uncorrelated\footnote{In a formal least-squares analysis, the $\epsilon_i$ terms must be independent and drawn from a zero-mean Gaussian distribution. We have relaxed the requirement that their distribution be Gaussian, and thus treat Equation \ref{eq:noise} as an approximation.}, then Equation \ref{eq:model} is an appropriate model, and a least-squares fit is expected to recover $a = \log \lambda + \overline{\log \left< N \right>}$ and $b=1$ to within some statistical uncertainty. This restriction on $\epsilon$ mandates that $\log \left<N\right>$ must not correlate with $A$. Equivalently, the mean of the column density PDF must not scale with cloud size for Equation \ref{eq:model} to be an appropriate model. If this is the case, then the statistical uncertainty on $b$ will be given by \citep[section 15.2]{Press07}
\begin{equation}
\sigma^2_b \sim \frac1{q} \frac{\sigma^2_{\log \left< N \right>}}{\sigma^2_{\log A}}
\label{eq:noise}
\end{equation}
Here $q$ is the sample size, and $\sigma^2_{\log \left< N \right>}$ and $\sigma^2_{\log A}$ are the sample variances of the $\log \left<N\right>_i$ and $\log A_i$ measurements, respectively. Thus, the precision at which L3 is recovered will be high when the dispersion of PDF means is small compared to the dispersion in areas. In summary, the conditions sufficient for a sample of mass and area measurements to follow L3 with low scatter are:
\begin{eqnarray}
\nonumber \left<N\right> \not \propto& A \\
\sigma_{\log \left<N\right>}^2 \ll & q ~ \sigma_{\log A}^2
\label{eq:necessary}
\end{eqnarray}

We use Equation \ref{eq:necessary} to determine when L3 is satisfied, and use these criteria as a more precise operational definition for cloud structure universality. These results provide three critiques to the analysis presented in \cite{Lombardi10}. First, as noted above, L3 is not specific to the log-normality of cloud column density distributions. Second, the slope of the mass-area relationship is influenced not only by the amount of total variation in the PDF, but also by any systematic scaling of the PDF with cloud size. Finally, in addition to the variation in PDF, both the sample size and dispersion in sizes influence the expected scatter about the $M \propto A$ relationship.

\begin{figure*}
\includegraphics[width=6in]{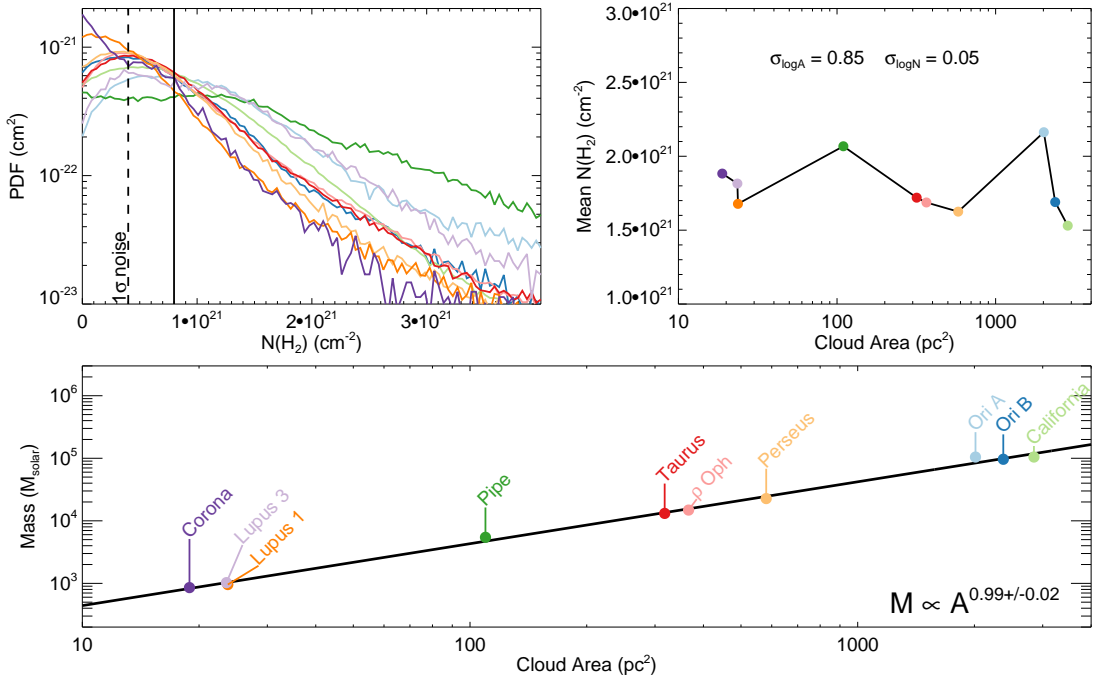}
\caption{The connection between the column density PDF and mass-area relationship for the data presented in \cite{Lombardi10}. Left: the column density PDF for each cloud. Right: the mean column density of pixels with extinctions higher than the threshold defined by the solid line on the left plot, as a function of cloud area. Bottom: The mass-area relationship.}
\label{fig:mr_data}
\end{figure*}

\begin{figure*}
\includegraphics[width=6in]{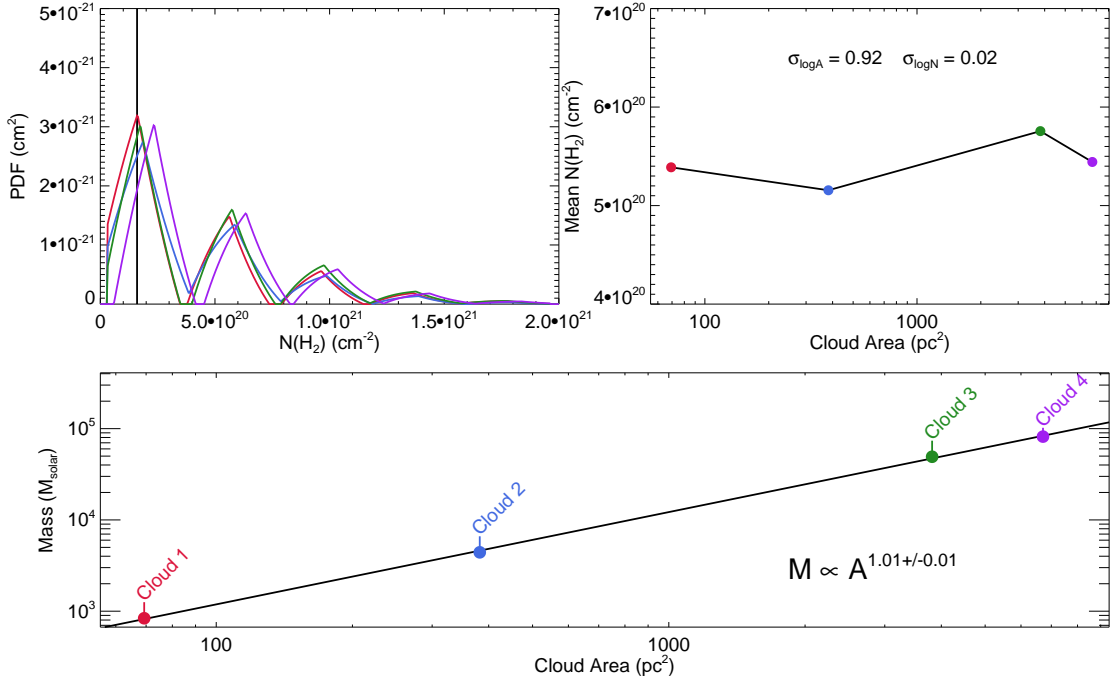}
\caption{The same as Figure \ref{fig:mr_data}, but for a fictional set of clouds. The PDFs of these regions are not log-normal, but still obey $M \propto A$.}
\label{fig:mr}
\end{figure*}

\begin{figure*}
\includegraphics[width=6in]{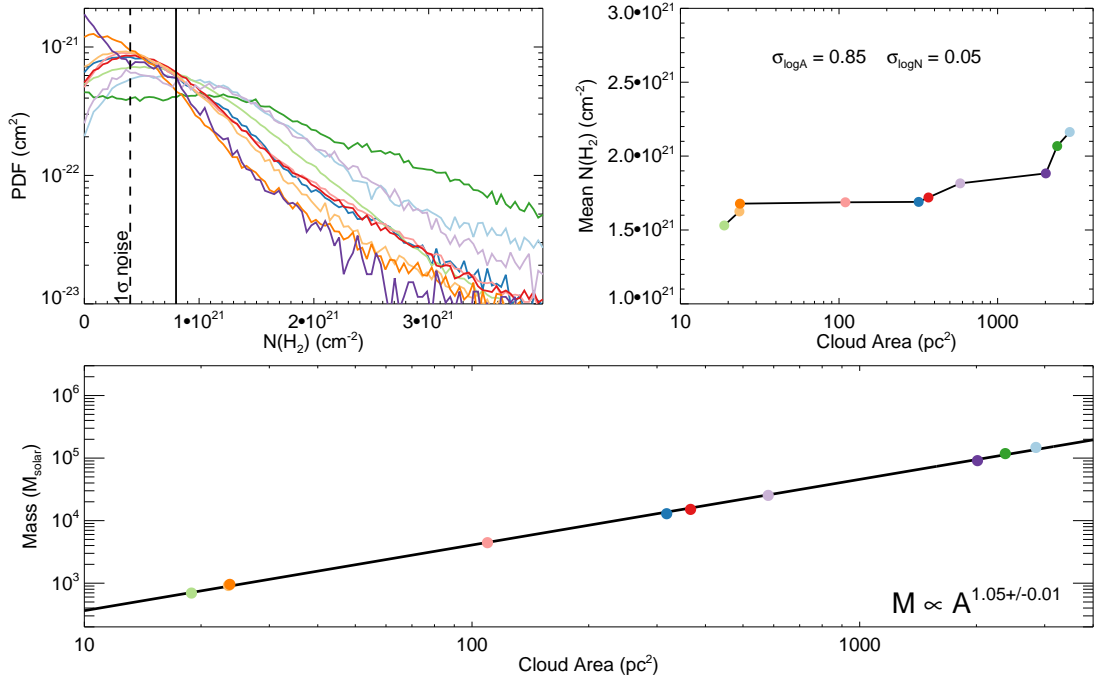}
\caption{The same as Figure \ref{fig:mr_data}, but with the cloud areas shuffled so as to correlate with mean column density. These clouds are inconsistent with L3 at the 5$\sigma$ level, even though the spread in mean column density and area is the same as in Figure \ref{fig:mr_data}.}
\label{fig:mr_correlate}
\end{figure*}

\begin{figure*}
\includegraphics[width=6in]{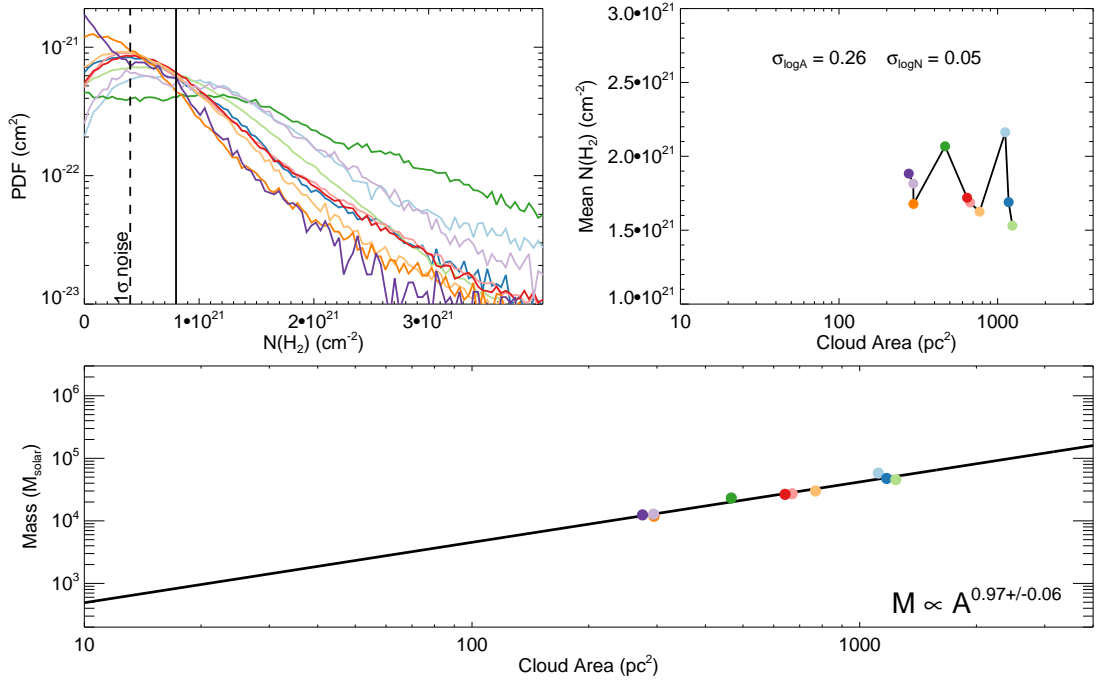}
\caption{The same as Figure \ref{fig:mr_data}, but with the dynamic range in areas artificially compressed. L3 is still an appropriate model, but the statistical errors are larger.}
\label{fig:mr_shrink}
\end{figure*}

These critiques are illustrated in Figures \ref{fig:mr_data}-\ref{fig:mr_shrink}. Each figure shows a collection of column density PDFs, the size and mean column density of each region, and the mass-area relationship. For reference, Figure \ref{fig:mr_data} shows the data presented in \cite{Lombardi10}. There are mild variations in the PDF for each cloud (left), but these variations introduce only minor scatter in the measurements of $\left<N\right>$, which are uncorrelated with size (right). Consequently, the clouds obey L3 (bottom). Figure \ref{fig:mr} shows an example of hypothetical structures with pathological, non-lognormal PDFs. These PDFs are also similar enough that $\left<N\right>$ is roughly constant, and these clouds also obey L3. Figure \ref{fig:mr_correlate} presents a modified version of the data in Figure \ref{fig:mr_data}. Here, the measurements of $A$ are shuffled so as to correlate with $\left<N\right>$. Thus, even though the scatter in $\left<N\right>$ and $A$ is the same as in Figure \ref{fig:mr_data}, Equation \ref{eq:necessary}a is violated, and the clouds are inconsistent with $M \propto A$ by $5 \sigma$. Finally, the dynamic range in areas for the data in Figure \ref{fig:mr_shrink} is artificially compressed, decreasing $\sigma_{\log A}$. While L3 is still an appropriate fit to the data, the scatter about the line is larger, as suggested by Equation \ref{eq:necessary}b.

\section{The intra-cloud mass-area relationship}
\label{sec:intra}
\begin{figure}
\includegraphics[width=3.5in]{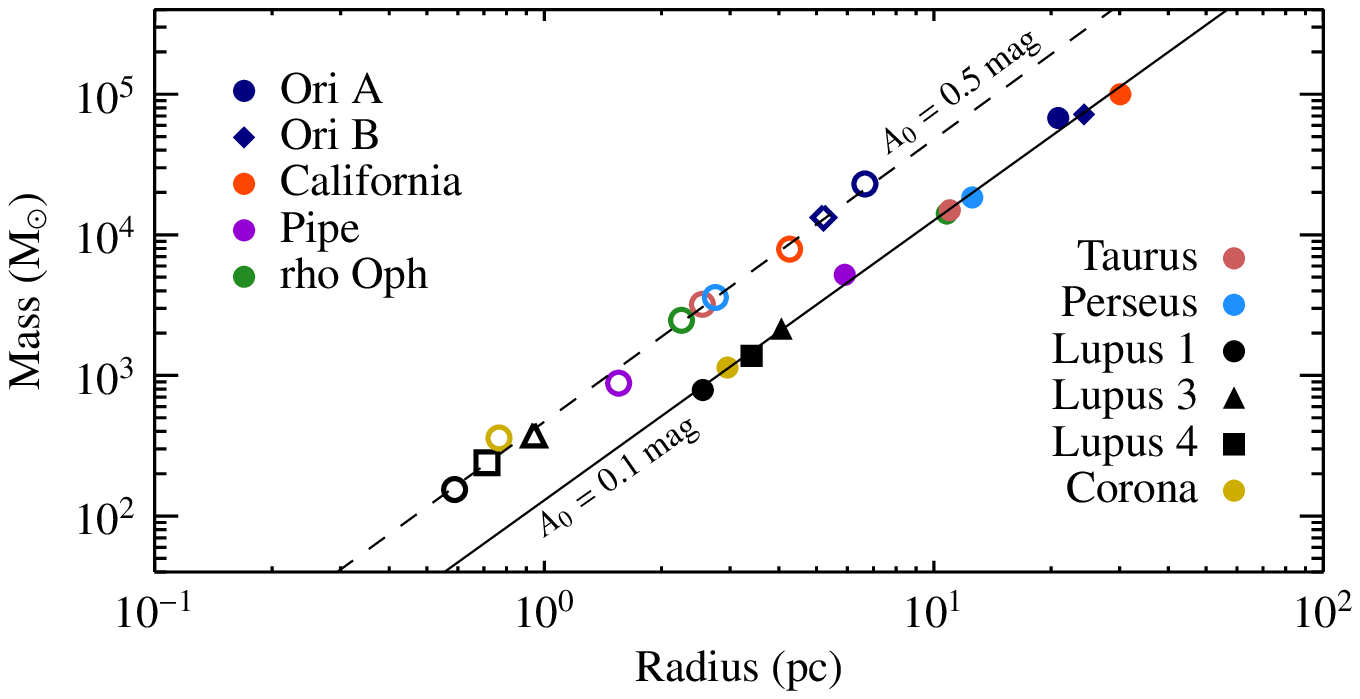}
\includegraphics[width=3in]{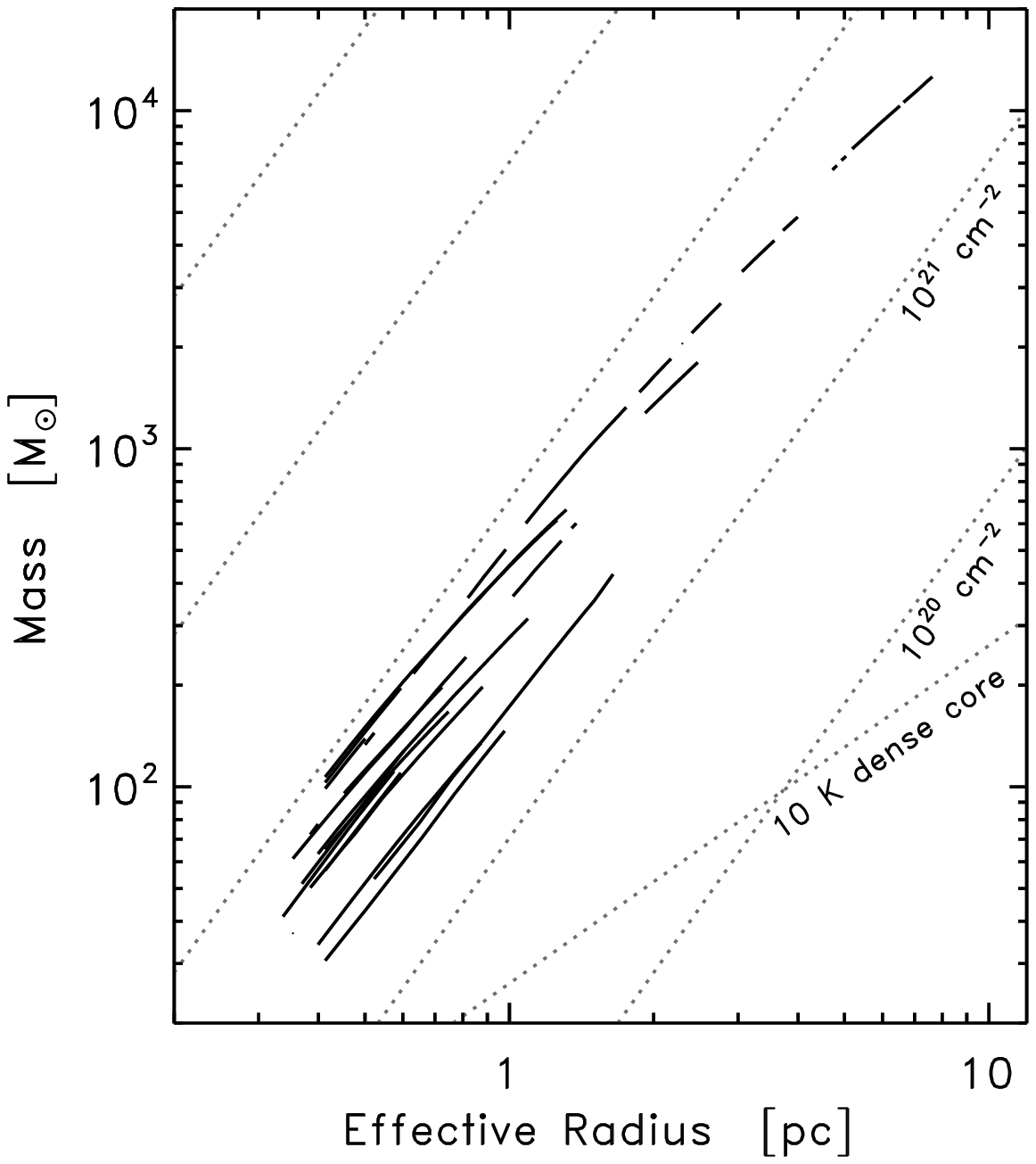}
\caption{Left: The inter-cloud mass-radius relationship, as published by \cite{Lombardi10}. The two lines correspond to cloud boundaries defined at to different extinction contours $A_0$(=$A_K$). The data obey $M \propto R^2$ to excellent approximation. Right: The intra-cloud mass-radius relationship for Perseus, as presented by \cite{Kauffmann10a}. Each line corresponds to a single substructure, sampled at different extinction contours. Different substructures appear to have significantly different relationships. Both figures define $R \equiv \sqrt{A / \pi}$}
\label{fig:compare}
\end{figure}

We are now in a position to address the mass-area relationship within a single cloud, and in particular the mass-area relationships presented by \cite{Kauffmann10a}. The main difference between the analyses in \cite{Kauffmann10a} and \cite{Lombardi10} is that, in the former work, each disconnected region within a cloud is treated separately, while such structures are aggregated in the latter. The mass-size plots presented in \cite{Kauffmann10a} exhibit larger scatter and deviations from $M \propto A$ (Figure \ref{fig:compare}b). Does this contradict the ``universality'' of cloud structure suggested by LAL10? Do cloud substructures exhibit substantial variation in the column density PDF that only average out when entire clouds are compared to one another?

The formalism discussed above can be applied to substructures as well -- in particular, the mass, area, and PDF for each substructure in Figure \ref{fig:compare}b are still related via Equations \ref{eq:mna} and \ref{eq:truncated_mean}, with the caveat that, in general, $P(x)$ can vary from region to region within a cloud. However, the lines drawn in Figure \ref{fig:compare}b trace substructures within Perseus at a variety of different contour levels -- they are an example of a Type 3 mass-area relationship presented in Figure \ref{fig:schematic_2}. Consequently, they conflate any regional change in the PDF with changes in $N_0$ (in other words, the value for $\left<N\right>$ in Equation \ref{eq:truncated_mean} depends on both $P(x)$ and $N_0$).

To more directly compare the internal mass-area relationship in Perseus to the inter-cloud relationship discussed above, we show in Figure \ref{fig:perseus} the mean column density as a function of substructure area at a variety of different thresholds. Each line adopts a fixed extinction threshold, and is thus an example of a Type 4 mass-area relationship. These plots possess the same features as those in Figure \ref{fig:mr_data}; in particular, changes in mean column density are uncorrelated with size, and $\sigma_{\log N} \ll \sigma_{\log A}$. In other words, the apparent scatter of the lines in Figure \ref{fig:compare}b is due to displaying several structures defined via different thresholds $N_0$, and the cloud substructure within Perseus is ``universal'' in the same sense as the inter-cloud comparison presented by LAL10.

\section{Discussion and Future Directions}
Recent studies of the mass-area relationship within and among molecular clouds can be understood from the perspective of a more fundamental quantity - the column density distribution function. Specifically, the mass-area relationship serves as a diagnostic for the degree to which the PDF varies within and among clouds. This has a certain observational appeal -- while many different techniques are used to measure cloud mass and size \citep{Sanchez05, Rosolowsky06, Heyer09, Kauffmann10a, Lombardi10}, the column density PDF is usually measured and characterized in a consistent way. Thus, different observational studies of the PDF are more easily inter-compared than studies of the mass-area relationship.

Larson's mass-area relationship is occasionally interpreted as an artifact of defining cloud boundaries at fixed column density thresholds -- the argument is that the mass of the region in this scenario is $M \sim N_{\rm thresh}\times A$. This relation is approximately valid for real clouds, given the fact that their column density PDFs are all similar and concentrated towards low column density. However, if column density PDFs \textit{did} vary substantially from region to region (as defined by Equation \ref{eq:necessary}), this would affect the masses, and would violate L3. Larson's mass-area relationship is fundamentally an assertion that the means of column density PDFs do not correlate with the sizes of cloud structures.

Volume and column density PDFs (\rhopdf, \npdf) have also received extensive theoretical attention. Log-normal \rhopdf s are often attributed to the action of isothermal, supersonic turbulence \citep{Vaz94, Padoan97, Passot98}. Various scenarios could map a log-normal \rhopdf to a log-normal \npdf. For example, if the typical line-of-sight cloud depth is smaller than the turbulence correlation length -- and relatively uniform across the cloud -- then the \npdf\, would simply be a scaled version of the \rhopdf, multiplied by the (constant) line-of-sight depth \citep{Vaz01, Ballesteros11}. Alternatively, if cloud depths are large compared to the correlation length, each line of sight samples many independent values of the \rhopdf. The column density along each line of sight converges to $\sim \left<\rho\right> \times L$, where $L$ is the cloud depth and $\left<\rho\right>$ is the mean volume density. If the distribution of cloud depths $L$ follows a log-normal, so will the \npdf. It seems plausible that real clouds lie somewhere between these extremes, with modulations in both cloud depth and volume density influencing the \npdf.

\begin{figure*}
\includegraphics[width=6in]{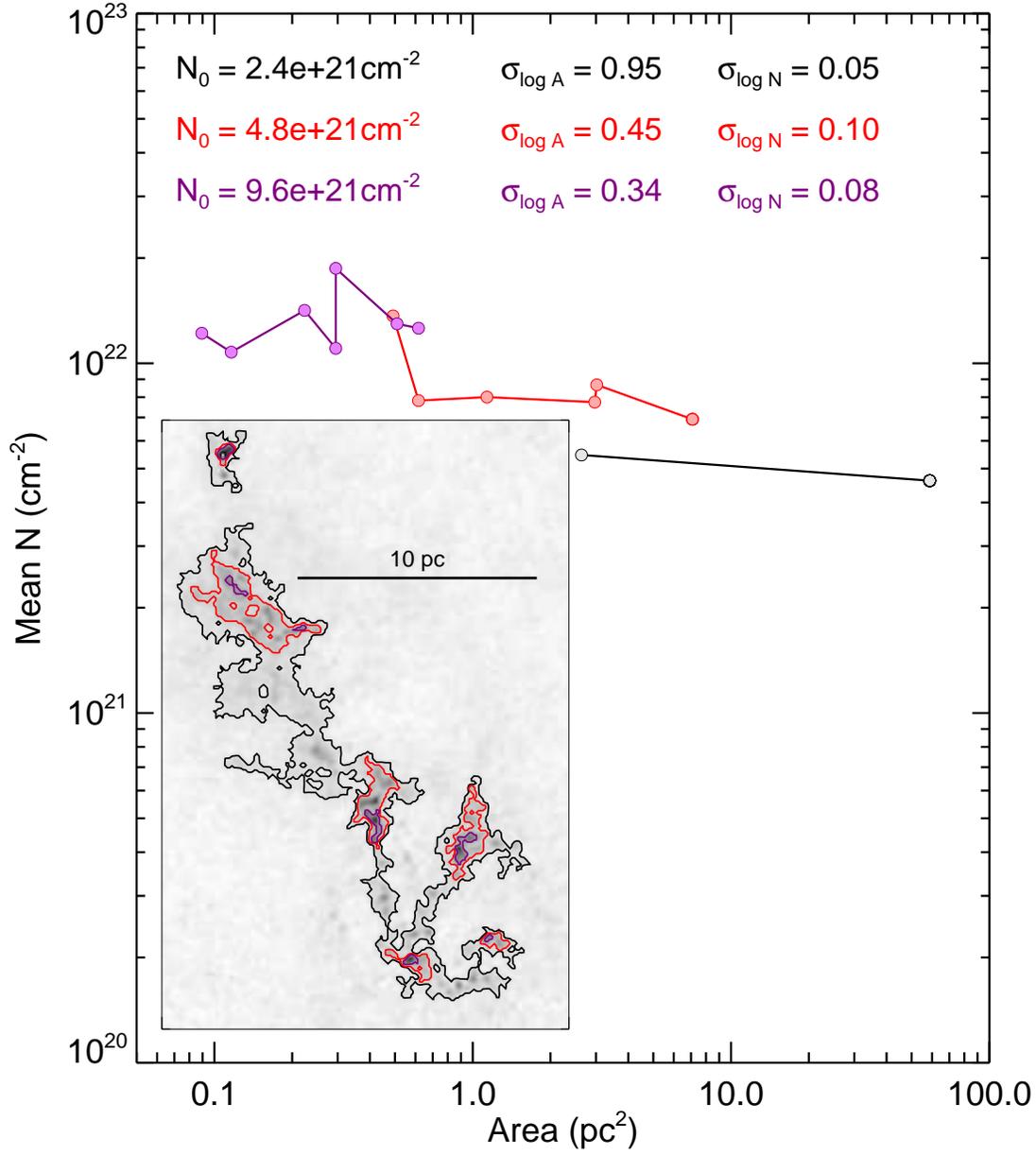}
\caption{The mean column density and area for substructures in the Perseus cloud, at three different column density thresholds. As in the inter-cloud relation, the spread in mean column density is small compared to the spread in area, and uncorrelated with area.}
\label{fig:perseus}
\end{figure*}

It has been suggested that the Mach number of supersonic turbulence determines the the width of the \rhopdf\, \citep{Passot98}. Since larger clouds exhibit larger velocity dispersions (Larson's first relationship), one might expect larger clouds to possess broader \npdf s and, by extension, larger values of $\left<N\right>$. Such a phenomenon would disobey L3, and is not seen in Figures \ref{fig:mr_data} or \ref{fig:perseus}.  Other processes may be acting to confine the width of cloud column density distributions. (e.g., magnetic turbulence, \citealt{Ostriker01}). \cite{Tassis10} has also demonstrated that several different physical processes besides supersonic turbulence can reproduce the column density PDFs observed for real clouds.

Gravity may act to create power-law excesses in the high column density tails of cloud PDFs. This view is supported observationally by the fact that non-star forming clouds tend to lack these tails \citep{Kainulainen09}, and that the star formation rate correlates with the integrated mass of the high-column density regions \citep{Lada10}. However, \cite{Alves12} suggest that these tails may also be consistent with the superposition of several cloud components, and \cite{Kainulainen11} have subsequently suggested that the high-column density tails are confined by pressure instead of gravity. Substructures within a cloud may well be dominated by different physical processes, inducing regional changes in the PDF. Figure \ref{fig:perseus} suggests that the mean of the PDF does not change dramatically within Perseus.

While the mass-area relationship can be viewed as a probe of the column density PDF and its variation, it contains only limited information about the PDF (namely, its mean). Given the amount of theoretical consideration of the \npdf, more carefully characterizing this distribution may provide more insight into what influences cloud structure. A rigorous observational study of the \npdf~ and its variations, however, must account for two factors:
\begin{enumerate}
\item Extinction measurements are still fairly noisy, and observed column density distributions convolve the true PDF with the measurement error kernel \citep{Kelly11}.
\item Clouds exhibit spatial correlations, and so neighboring measurements of the column density do not constitute independent samples of the underlying \npdf. Fitting procedures which do not account for this will over-estimate the effective number of samples, and under-estimate uncertainties in the fit to the PDF.
\end{enumerate}
To date, most characterizations of cloud PDFs have ignored these issues. However, they must be addressed to properly constrain the shape and spatial variation of the PDF.

\section{Conclusion}
We have directly compared the mass-area relationship within and among clouds, interpreting this relationship from the perspective of the column density distribution and its spatial variation. For structures defined via a fixed extinction contour, the mean of the column density PDF in each region varies less than, and is uncorrelated with, the region-to-region dispersion in area. This naturally yields the relationship $M \propto A$, and thus suggests that Larson's third scaling relationship holds both within and among clouds.

The structure of nearby molecular clouds is ``universal'' in the sense that the dispersion of the PDF mean is negligible compared to the dispersion of cloud areas. However, cloud structure may still vary in interesting ways that do not greatly affect the mean of the \npdf. We suggest that studying the shape and variation of the \npdf\, directly may provide more insight into the universality of cloud structure.

This material is based upon work supported by the National Science Foundation under Grant No. AST-0908159. CRZ acknowledges support from program UNAM-DGAPA-PAPIIT IA101812-2.

\begin{deluxetable}{rrrll}
\tablewidth{0in}
\tablecaption{Literature Fits to $(M/M_\odot) = \gamma (A/{\rm pc}^2)^\beta$}
\tablehead{
\colhead{$\gamma$} &
\colhead{$\beta$} &
\colhead{Reference} &
\colhead{Tracer} &
\colhead{Type\tablenotemark{a}}}
\startdata
240 & 0.95 & \cite{Larson81} & $^{13}$CO, $^{12}$CO, NH$_3$ & 1\\
170 & 1 & \cite{Solomon87} & $^{12}$CO & 1 \\
30 & 1-1.2 & \cite{Falgarone92} & $^{12}$CO, $^{13}$CO & 1 \\
50-500 & 1.2-1.8 & \cite{Elmegreen96} & $^{12}$ CO, $^{13}$CO, C$^{18}$O & 1\\
--- & 0.91 & \cite{Sanchez05} & $^{13}$CO & 3 \\
150 & 1.3 & \cite{Lada08} & Extinction & 4 \\
42 & 1 & \cite{Heyer09} & $^{13}$CO & 1 \\
228 & 1.36 & \cite{Roman10} & $^{12}$CO, $^{13}$CO & 1 \\
70 & 0.5---1 & \cite{Kauffmann10a} & Extinction, mm-continuum & 3 \\
41--380 & 0.99 -- 1.01& \cite{Lombardi10} & Extinction & 2\\
150 & 1.3 & \cite{RomanZ10} & Extinction & 4\\
\enddata
\tablenotetext{a}{Mass-area comparison type, as defined by Figure \ref{fig:schematic_2}.}
\label{tab:larson3}
\end{deluxetable}

\end{document}